\documentstyle [12pt] {article}

\catcode`@=12
\begin{document}
\thispagestyle{empty}
\begin{flushright} UCRHEP-T149\\July 1995\
\end{flushright}
\vspace{0.5in}
\begin{center}
{\large \bf Neutrino Masses in an Extended Gauge Model\\
with E$_6$ Particle Content\\}
\vspace{1.5in}
{\bf Ernest Ma\\}
{\sl Department of Physics\\}
{\sl University of California\\}
{\sl Riverside, California 92521\\}
\vspace{1.5in}
\end{center}
\begin{abstract}\
Naturally light singlet neutrinos which mix with the usual doublet neutrinos
are possible if the supersymmetric standard gauge model is extended to
include a specific additional $U(1)$ factor derivable from an $E_6$
decomposition.  The low-energy particle content of the model is limited to
the fundamental $\bf 27$ representations of $E_6$.
\end{abstract}

\newpage
\baselineskip 24pt

The three known neutrinos $\nu_e$, $\nu_\mu$, and $\nu_\tau$ are each a
component of an $SU(2) \times U(1)$ doublet, pairing with the left-handed
projections of the charged leptons $e$, $\mu$, and $\tau$ respectively.
They are generally considered to be Majorana fermions with very small masses
arising from the so-called ``seesaw" mechanism.\cite{1}  This means that there
should be three heavy neutral fermion singlets $N_{1,2,3}$ which also couple
to $\nu_{e,\mu,\tau}$ through the usual Higgs doublet $\Phi = (\phi^+,\phi^0)$
of the standard model.  As $\phi^0$ acquires a nonzero $vev$ (vacuum
expectation value), a Dirac mass term $m_D$ linking $\nu$ and $N$ is obtained,
yielding the well-known result $m_\nu \simeq m_D^2/m_N$.  The most natural
origin of $N_{1,2,3}$ is that associated with a left-right model where they
can be identified as the right-handed counterparts of the left-handed
neutrinos. As the $SU(2)_L \times SU(2)_R \times U(1)$ gauge symmetry breaks
down to the standard $SU(2) \times U(1)$, a large $m_N$ may be obtained.

Whereas the usual seesaw mechanism is based on a $2 \times 2$ matrix
\begin{equation}
{\cal M}_2 = \left( \begin{array} {c@{\quad}c} 0 & m_D \\ m_D & m_N
\end{array} \right)
\end{equation}
with the doublet neutrino getting a small mass, there is also the
simple variation where it is the \underline {singlet} neutrino which gets
a small mass.  Consider the left-handed fermion doublets $(\nu_E, E)$ and
$(E^c, N_E^c)$ transforming as $(2,-1/2)$ and (2,1/2) respectively under
the standard $SU(2) \times U(1)$.  Add a neutral fermion singlet $S$ and
forbid it to have a Majorana mass term by an appropriate symmetry.
The $3 \times 3$ mass matrix spanning $\nu_E$, $N_E^c$, and $S$ may then
be given by
\begin{equation}
{\cal M}_3 = \left( \begin{array} {c@{\quad}c@{\quad}c} 0 & m_E & m_1 \\
m_E & 0 & m_2 \\ m_1 & m_2 & 0 \end{array} \right),
\end{equation}
where $m_{1,2}$ are proportional to the $vev$ of an appropriate Higgs
doublet and $m_E$ is
now an allowed gauge-invariant mass.  For $m_{1,2} << m_E$, we then have
$m_S \simeq 2 m_1 m_2/m_E$.  If ${\cal M}_3$ is also linked to ${\cal M}_2$,
then the light singlet $S$ will also mix with the usual doublet neutrinos.

If a light singlet neutrino exists in addition to the three doublet
neutrinos, a comprehensive picture of neutrino oscillations and hot dark
matter becomes possible.\cite{2}  This is especially so because of the
recent results of the LSND (Liquid Scintillator Neutrino Detector)
experiment\cite{3} which may be interpreted as $\nu_\mu$ oscillating to
$\nu_e$ with a $\Delta m^2$ of a few eV$^2$.  To avoid the severe
constraint on the effective number of neutrinos from big-bang
nucleosynthesis,\cite{4} the singlet neutrino may be used only to account
for the solar data by mixing with $\nu_e$ in the matter-enhanced
small-angle solution or the long-wavelength large-angle solution.

A good model for a light singlet neutrino should have an appropriate
symmetry which forbids it to have a Majorana
mass term, as already noted.  It is of course easy to impose such a
symmetry, but for it to be natural, it should come from a more fundamental
framework, such as grand unification or string theory for example.  As it
turns out, Eq.~(2) is a natural consequence of the superstring-inspired
$E_6$ model.\cite{5}  Unfortunately, the corresponding $m_N$ of Eq.~(1)
is zero there.\cite{6}  This means that $\nu_{e,\mu,\tau}$ combine with
$N_{1,2,3}$ to form Dirac neutrinos and their small masses are unexplained.
On the other hand, gravitationally induced nonrenormalizable
interactions\cite{7}
may produce large Majorana mass terms for both $N$ and $S$, in
which case $\nu_{e,\mu,\tau}$ are again naturally light by virtue of the
seesaw mechanism, but they will be the only ones.

The low-energy gauge symmetry of a superstring-inspired $E_6$ model is
often taken to be $SU(3)_C \times SU(2)_L \times U(1)_Y \times U(1)_\eta$,
because the flux mechanism of symmetry breaking in string theory involves
the adjoint representation and it is not possible\cite{5} to break $E_6$
all the way down to the gauge symmetry of the standard model.  If only
one extra $U(1)$ factor is present, then it is necessarily $U(1)_\eta$,
according to which both $N$ and $S$
transform nontrivially.  They are thus protected by this gauge symmetry
from acquiring large Majorana masses.  For the nonrenormalizable mechanism
of Ref.~[7] to work, the $U(1)_\eta$ must also be broken at an
intermediate scale by $vev$'s along the $N$ and $S$ directions.
To obtain a light neutrino doublet with Eq.~(1) as well as a light neutrino
singlet with Eq.~(2), the idea then is to replace $U(1)_\eta$ with
another $U(1)$ under which $N$ is trivial but $S$ is not, so that
only the former may acquire a large Majorana mass.  In the following
this extended gauge model is described.

Consider the maximal subgroup $SU(3)_C \times SU(3)_L \times SU(3)_R$ of
$E_6$.  The fundamental {\bf 27} representation of $E_6$ is then given by
\begin{equation}
{\bf 27} = (3,3,1) + (3^*,1,3^*) + (1,3^*,3).
\end{equation}
Under the decompositions $SU(3)_L \rightarrow SU(2)_L \times U(1)_{Y_L}$
and $SU(3)_R \rightarrow U(1)_{T_{3R}} \times U(1)_{Y_R}$, the
individual left-handed fermionic components are defined as follows.\cite{8}
\begin{eqnarray}
(u,d) &\sim& (3;2,{1 \over 6};0,0), \\ (\nu_e,e) &\sim& (1;2,-{1 \over 6};
0,-{1 \over 3}), \\ u^c &\sim& (3^*;0,0;-{1 \over 2},-{1 \over 6}), \\
d^c &\sim& (3^*;0,0;{1 \over 2},-{1 \over 6}), \\ e^c &\sim& (1;0,{1 \over 3};
{1 \over 2},{1 \over 6}), \\ N &\sim& (1;0,{1 \over 3};-{1 \over 2},
{1 \over 6}), \\ h &\sim& (3;0,-{1 \over 3};0,0), \\ h^c &\sim& (3^*;0,0;
0,{1 \over 3}), \\ (\nu_E,E) &\sim& (1;2,-{1 \over 6};-{1 \over 2},
{1 \over 6}), \\ (E^c,N_E^c) &\sim& (1;2,-{1 \over 6};{1 \over 2},
{1 \over 6}), \\ S &\sim& (1;0,{1 \over 3};0,-{1 \over 3}).
\end{eqnarray}
Note that the electric charge is given here by
\begin{equation}
Q = T_{3L} + Y_L + T_{3R} + Y_R,
\end{equation}
and there are three families of these fermions and their bosonic
superpartners.

Consider now the $SO(10)$ decomposition of the {\bf 27} representation:
\begin{equation}
{\bf 27} = {\bf 16} + {\bf 10} + {\bf 1}.
\end{equation}
Two options are available.  The conventional one (Option A) is
\begin{eqnarray}
{\bf 16} &=& (u,d) + u^c + e^c + d^c + (\nu_e,e) + N, \\ {\bf 10} &=&
h + (E^c,N_E^c) + h^c + (\nu_E,E), \\ {\bf 1} &=& S.
\end{eqnarray}
The alternative one (Option B) is\cite{9}
\begin{eqnarray}
{\bf 16} &=& (u,d) + u^c + e^c + h^c + (\nu_E,E) + S, \\ {\bf 10} &=&
h + (E^c,N_E^c) + d^c + (\nu_e,e), \\ {\bf 1} &=& N.
\end{eqnarray}
The latter is obtained from the former by the exchange\cite{9}
\begin{equation}
d^c \leftrightarrow h^c, ~~~ (\nu_e,e) \leftrightarrow (\nu_E,E), ~~~
N \leftrightarrow S,
\end{equation}
so that $SU(3)_R$ is broken along a different direction, namely that given by
\begin{equation}
T'_{3R} = {1 \over 2} T_{3R} + {3 \over 2} Y_R, ~~~ Y'_R = {1 \over 2}
T_{3R} - {1 \over 2} Y_R.
\end{equation}
As far as the standard $SU(3)_C \times SU(2)_L \times U(1)_Y$ gauge
symmetry is concerned, the two options are identical because
\begin{equation}
T'_{3R} + Y'_R = T_{3R} + Y_R = Q - T_{3L} - Y_L.
\end{equation}
In the $U(1)_\eta$ extension, it can also be shown that there is no
difference because $Q_\eta$ is proportional to $T_{3L} + 5Y_L - Q$.\cite{8}
Furthermore, the same Yukawa terms are allowed by either option,
independent of any additional $U(1)$.  This is easily seen by expressing
the {\bf 27} representation in terms of its ($SO(10)$, $SU(5)$) components:
\begin{equation}
{\bf 27} = ({\bf 16},{\bf 10}) + ({\bf 16}, {\bf 5^*}) + ({\bf 16}, {\bf 1})
+ ({\bf 10}, {\bf 5}) + ({\bf 10}, {\bf 5^*}) + ({\bf 1}, {\bf 1}).
\end{equation}
The allowed terms must then be of the form
({\bf 16},{\bf 10})({\bf 16},{\bf 10})({\bf 10},{\bf 5}),
({\bf 16},{\bf 10})({\bf 16},{\bf 5$^*$})({\bf 10},{\bf 5$^*$}),
({\bf 10},{\bf 5})({\bf 16},{\bf 5$^*$})({\bf 16},{\bf 1}), and
({\bf 10},{\bf 5})({\bf 10},{\bf 5$^*$})({\bf 1},{\bf 1}), which remain the
same if ({\bf 16},{\bf 5$^*$}) and ({\bf 10},{\bf 5$^*$}) are exchanged
together with ({\bf 16},{\bf 1}) and ({\bf 1},{\bf 1}), in accordance with
Eq.~(23).

Two $U(1)$ factors are conventionally defined in Option A by the symmetry
breaking chain
\begin{equation}
E_6 \rightarrow SO(10) \times U(1)_\psi, ~~~ SO(10) \rightarrow SU(5)
\times U(1)_\chi.
\end{equation}
If the extended gauge model contains only one additional $U(1)$ factor,
it must be a linear combination of $U(1)_\psi$ and $U(1)_\chi$.  Let
\begin{equation}
Q(\alpha) = Q_\psi \cos \alpha + Q_\chi \sin \alpha,
\end{equation}
then the $U(1)_\eta$ from flux breaking corresponds to $\tan \alpha =
\sqrt {3/5}$.  On the other hand, the $U(1)$ factor for which $N$ is
trivial is clearly that which would be called $U(1)_\chi$ in Option B.
This turns out to be given by $\tan \alpha = -\sqrt {1/15}$.  To obtain
this factor which will be called $U(1)_N$ from now on, the flux breaking
of $E_6$ must be augmented by the usual Higgs mechanism, presumably at
near the same scale.  Consider then a pair of superheavy {\bf 27} and
{\bf 27$^*$} representations.  Assume that they develop $vev$'s along
the $N$ and $N^*$ directions respectively.  Then $E_6$ is broken down to
the $SO(10)$ of Option B.  Assume also that the flux mechanism breaks
$SU(3)_L$ to $SU(2)_L \times U(1)_{Y_L}$, {\it i.e.} along the (1,0)
direction, and $SU(3)_R$ to $U(1)_{T_{3R}} \times U(1)_{Y_R}$, {\it i.e.}
along the (3,0) direction.  Then the resulting gauge symmetry is exactly
$SU(3)_C \times SU(2)_L \times U(1)_Y \times U(1)_N$ with
\begin{equation}
Q_N = 6 Y_L + T_{3R} - 9 Y_R.
\end{equation}
The individual particle assignments under $U(1)_N$ are then
\begin{eqnarray}
(u,d), u^c, e^c &:& 1, \\ d^c, (\nu_e, e) &:& 2, \\ h, (E^c, N_E^c) &:& -2, \\
h^c, (\nu_E, E) &:& -3, \\ S &:& 5, \\ N &:& 0.
\end{eqnarray}
As in any other superstring-inspired $E_6$ model, a discrete symmetry must
be imposed to eliminate rapid proton decay.\cite{10}  Here a $Z_2$ symmetry
is assumed where all superfields are odd except one copy each of $(\nu_E,E)$,
$(E^c,N_E^c)$, and $S$, which are even.
Consequently, the allowed cubic terms of the superpotential are
$u^c (u N_E^c - d E^c)$, $d^c (u E - d \nu_E)$, $e^c (\nu_e E - e \nu_E)$,
$S (E E^c - \nu_E N_E^c)$, $S h h^c$, and
$N (\nu_e N_E^c - e E^c)$.  As the scalar components of the even superfields
$\nu_E$, $N_E^c$, and $S$ acquire $vev$'s, all particles obtain masses
in the usual way.  In addition, since $N$ is now a gauge singlet, it may
acquire a large Majorana mass from nonrenormalizable interactions.\cite{7}
The quadratic terms $d^c h$ and $\nu_e N_E^c - e E^c$ are also gauge
singlets, and allowed by the discrete $Z_2$ symmetry.  [The latter term is
of course restricted to the two odd $(E^c,N_E^c)$ doublets.]  They are soft
terms which reduce the symmetries of the Lagrangian and may thus be assumed
to be naturally small.\cite{11}  Their origin is presumably also from
nonrenormalizable interactions.  Note that both baryon number and lepton
number remain conserved.

Consider now the $5 \times 5$ mass matrix spanning $\nu_e$, $N$, $\nu_E$,
$N_E^c$, and $S$.  It is exactly given by combining Eq.~(1) with Eq.~(2)
and adding a $\nu_e N_E^c$ term:
\begin{equation}
{\cal M}_5 = \left( \begin{array} {c@{\quad}c@{\quad}c@{\quad}c@{\quad}c}
0 & m_D & 0 & m_3 & 0 \\ m_D & m_N & 0 & 0 & 0 \\ 0 & 0 & 0 & m_E & m_1 \\
m_3 & 0 & m_E & 0 & m_2 \\ 0 & 0 & m_1 & m_2 & 0 \end{array} \right),
\end{equation}
where $m_D$ and $m_1$ come from $\langle \tilde N_E^c \rangle$, $m_2$ from
$\langle \tilde \nu_E \rangle$, and $m_E$ from $\langle \tilde S \rangle$.
The $S$ fermion corresonding to the last $vev$ is even under $Z_2$ and
it becomes massive because $U(1)_N$ is broken by $\langle \tilde S
\rangle$ through which a mass term is generated, linking it with the
corresponding gauge fermion.  The two odd $S$'s remain light and are naturally
suited to be light singlet neutrinos.  For illustration, let $m_1 = m_2
= m_e = 0.5$ MeV and $m_E = 2 \times 10^5$ GeV, then $m_S \simeq 2 m_1 m_2/m_E
= 2.5 \times 10^{-3}$ eV.  Assuming that $m_{\nu_e}$ is much smaller, then
$\nu_e - \nu_S$ oscillations occur with $\Delta m^2 \simeq 6 \times 10^{-6}~
{\rm eV}^2$ which is in the right range to account for the solar data.
The mixing angle between $\nu_e$ and $\nu_S$ is given by $m_3/2m_2$ and
should be about 0.04 for $\sin^2 2 \theta \simeq 6 \times 10^{-3}$.

In conclusion, a supersymmetric extended gauge model based on $SU(3)_C
\times SU(2)_L \times U(1)_Y \times U(1)_N$ has been proposed.  This
gauge symmetry is derivable from an $E_6$ superstring model through a
combination of flux breaking and the usual Higgs mechanism with a pair
of superheavy {\bf 27} and {\bf 27$^*$} representations.  Its particle
content consists of three supermultiplets belonging to the fundamental
{\bf 27} representation of $E_6$ as listed in Eqs.~(4) to (14) and $Q_N$ is
given by Eq.~(29).  The three $N$ singlets are trivial under $U(1)_N$ and
naturally acquire large Majorana masses from gravitationally induced
nonrenormalizable interactions.  One of the $S$ singlets has a $vev$
which breaks $U(1)_N$ at an unspecified scale and renders all remaining
particles heavy except for those of the supersymmetric standard model and
the other two $S$ singlets.  At and below the electroweak energy scale,
this model differs from the minimal supersymmetric standard model (MSSM)
in the following important ways.  (1) The three known doublet neutrinos
$\nu_e$, $\nu_\mu$, and $\nu_\tau$ have small Majorana masses instead of
being massless as in the MSSM.  (2) Two light singlet neutrinos exist and
they may have small mixings with $\nu_e$, $\nu_\mu$, and $\nu_\tau$.
This allows for a comprehensive understanding of neutrino oscillations
as well as hot dark matter in the face of all available data.
(3) The scalar partners of one set of the $(\nu_E,E)$ and $(E^c,N_E^c)$
superfields are identified with the two usual Higgs doublets $\Phi_1$
and $\Phi_2$ of the MSSM.  However, the Higgs potentials are different
because the superpotential here has the cubic term $(\nu_E N_E^c - E E^c)S$
whereas the MSSM has the quadratic term $\phi_1^0 \phi_2^0 - \phi_1^-
\phi_2^+$.  Hence the corresponding higgsino mass is bounded here by
$\langle \tilde S \rangle$ whereas in the MSSM, there is no understanding
as to why this mass should be much smaller than the unification scale
of 10$^{16}$ GeV or the Planck scale of 10$^{19}$ GeV.  The Higgs potential
of this model has only two doublets at the electroweak energy scale, but
because of the above-mentioned cubic term in the superpotential, it differs
from that of the MSSM by one extra coupling.  Previous such examples have
been given for other gauge extensions.\cite{12}
\vspace{0.3in}
\begin{center} {ACKNOWLEDGEMENT}
\end{center}

I thank A. Mendez for hospitality and a discussion at the Univ. Autonoma
de Barcelona which led to this investigation.  This
work was supported in part by the U. S. Department of Energy under Grant
No. DE-FG03-94ER40837.

\bibliographystyle{unsrt}

\end{document}